\documentclass[conference]{IEEEtran}
\IEEEoverridecommandlockouts
\usepackage{cite}
\usepackage{amsmath,amssymb,amsfonts}
\usepackage{algorithmic}
\usepackage{graphicx}
\usepackage{textcomp}
\usepackage{xcolor}
\usepackage[a4paper, total={184mm,239mm}]{geometry}
\def\BibTeX{{\rm B\kern-.05em{\sc i\kern-.025em b}\kern-.08em
    T\kern-.1667em\lower.7ex\hbox{E}\kern-.125emX}}
\usepackage{filecontents}
\usepackage{tikz}
\newcommand*\circled[1]{\tikz[baseline=(char.base)]{
            \node[shape=circle,draw,inner sep=1pt] (char) {#1};}}

\newcommand{\covert}{\textsf{{COVERT}}}

\usepackage{url}       
\usepackage[T1]{fontenc}    
\usepackage{underscore}     

\usepackage{tabularx,array,ragged2e,url} 
\newcolumntype{Y}{>{\RaggedRight\arraybackslash}X}          
\newcolumntype{Z}{>{\ttfamily\RaggedRight\arraybackslash}X} 

\usepackage{fancyhdr}
\fancypagestyle{firstpage}{

\fancyfoot[C]{\large{\textcolor{blue}{This paper has been accepted for publication at DATE 2026. This is an initial author version.\\ The final paper will appear in the conference proceedings.}}}

}

\begin{document}

\title{COVERT: Trojan Detection in COTS Hardware via Statistical Activation of Microarchitectural Events}

\author{
\IEEEauthorblockN{Mahmudul Hasan\IEEEauthorrefmark{1}, Sudipta Paria\IEEEauthorrefmark{2}, Swarup Bhunia\IEEEauthorrefmark{2} and Tamzidul Hoque\IEEEauthorrefmark{1}}
\IEEEauthorblockA{\IEEEauthorrefmark{1}Department of Electrical Engineering and Computer Science, University of Kansas, Lawrence, KS 66045, USA\\
\IEEEauthorrefmark{2}Department of Electrical and Computer Engineering, University of Florida, Gainesville, FL 32611, USA\\
Email: \{m.hasan@ku.edu, sudiptaparia@ufl.edu, swarup@ece.ufl.edu, hoque@ku.edu\}}
}

\maketitle
\thispagestyle{firstpage}

\begin{abstract}
Commercial Off-The-Shelf (COTS) hardware, such as microprocessors, are widely adopted in system design due to their ability to reduce development time and cost compared to custom solutions. However, supply chain entities involved in the design and fabrication of COTS components are considered untrusted from the consumer’s standpoint due to the potential insertion of hidden malicious logic or hardware Trojans (HTs). Existing solutions to detect Trojans are largely inapplicable for COTS components due to their black-box nature and lack of access to a golden model. A few studies that apply require expensive equipment, lack scalability, and apply to a limited class of Trojans. In this work, we present a novel golden-free trust verification framework, \covert,~for COTS microprocessors, which can efficiently test the presence of hardware Trojan implants by identifying microarchitectural rare events and transferring activation knowledge from existing processor designs to trigger highly susceptible internal nodes. \covert~leverages Large Language Models to automatically generate test programs that trigger rare microarchitectural events, which may be exploited to develop Trojan trigger conditions. By deriving these events from publicly available Register Transfer Level implementations, \covert~can verify a wide variety of COTS microprocessors that inherit the same Instruction Set Architecture. We have evaluated the proposed framework on open-source RISC-V COTS microprocessors and demonstrated its effectiveness in activating combinational and sequential Trojan triggers with high coverage, highlighting the efficiency of the trust verification. 
By pruning rare microarchitectural events from mor1kx Cappuccino OpenRISC processor design, \covert~has been able to achieve more than 80\% trigger coverage for the rarest 5\% of events in or1k Marocchino and PicoRV32 as COTS processors.
\end{abstract}

\begin{IEEEkeywords} 
Trojan Detection,
Test Generation,
Microarchitectural Events,
COTS,
Combinational and Sequential Trojans.

\end{IEEEkeywords}

\section{Introduction}

Commercial off-the-shelf (COTS) components offer a compelling system design paradigm due to reduced development time, lower hardware costs, and market availability compared to custom design approach. As a result, COTS microprocessors have been widely adopted in military, avionics, finance, and commercial sectors. According to a 2022 report, around 98\% of microelectronic components used in defense applications are COTS~\cite{shivakumar2022semiconductors}. The increasing reliance on COTS components has simultaneously introduced significant security concerns. Suppliers of COTS components distribute design, manufacturing, and testing across various domestic and foreign untrusted entities. Any untrusted entities with access to the design could introduce hidden malicious logic or hardware Trojans capable of causing functional failures or leakage of sensitive information such as encryption keys \cite{xiao2016hardware}. Many real-world cyber attacks indicate the rising threat of hardware Trojans in untrusted components \cite{sarker2025everyday}.

Most existing hardware Trojan research primarily focuses on insertion threats either from untrusted foundry or untrusted IP vendor \cite{jain2021survey, cruz2022automatic}. Most of these countermeasures rely on one or more of the following: (i) access to golden (Trojan-free) chips, (ii) white-box access to the design, or (iii) design-time modifications to integrate countermeasures \cite{xiao2016hardware}. None of these assumptions hold for COTS components, rendering existing methods ineffective. For example, logic testing approaches depend on design access to generate effective test vectors \cite{chakraborty2009mero,latent}, while most side-channel analysis (SCA) methods require golden chips to establish reference signatures and design access for test generation to ensure switching activity in Trojans \cite{8353873}.
A few studies have developed runtime approaches applicable to COTS hardware to detect, tolerate, or prevent Trojan activation in the field \cite{variant_cots, cassano2022deton, ref:safer}. While runtime solutions introduce an additional layer of security, trust verification approach is generally more desirable as the primary defense mechanism. Side-channel assisted golden-chip free trust-verification methods have been developed that can be applied to COTS hardware, but these techniques also assume the presence of design-aware test-generation methods that can ensure activation of Trojan nets \cite{stern2020sparta, yang2021trusted}. Therefore, there is a critical unmet need for high-confidence black-box trust verification of COTS hardware.      

In this paper, we propose \covert~(\underline{CO}TS \underline{VER}ification Framework for \underline{T}rojan Detection), a novel test generation approach for hardware Trojan detection in COTS microprocessors through statistical activation of microarchitectural events.
\covert~leverages the software code generation capabilities of Large Language Models (LLMs) to generate targeted test programs for activating the rare internal events that can form the triggers of potential Trojans. While the golden design of a COTS processor may not be available, a procurer of the COTS component typically receives comprehensive documentation of the instruction set architecture (ISA) and architectural features, including the list of interrupts, events, and limited hardware components details. \covert~leverages this publicly available ISA together with an open-source RTL implementation of that ISA to explore rare microarchitectural events. The majority of these events are implementation agnostic, meaning they are applicable to COTS processors of the same ISA.  Next, \covert~construct targeted test programs that exercise those rare microarchitectural events, which are common targets for crafting hard-to-activate Trojan trigger conditions. \covert~leverages the relationship across design abstractions to connect low-level Trojan nets with high-level events.

\begin{figure}[hbt]
    \centering
    \setlength{\belowcaptionskip}{-30pt}
    \includegraphics[width=\columnwidth]{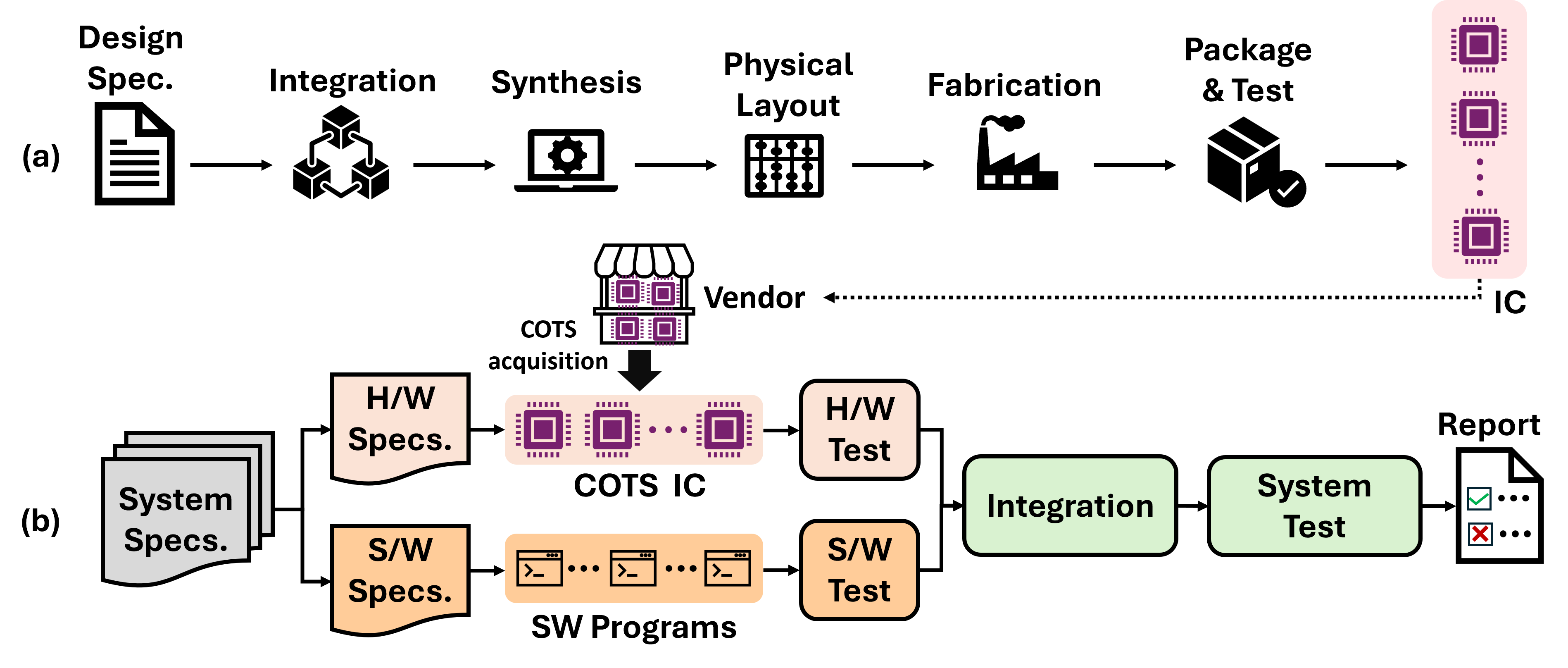}
    \vspace{-1em}
     \caption{(a) Modern IC design flow; (b) COTS IC integration flow \cite{cots_survey,ip_cots}.}
    \label{fig:design_flow}
\end{figure}

Microarchitectural events, such as pipeline hazards, cache behaviors, or exception conditions are directly derived from the RTL design and represent exact functional scenarios within the processor. If an event is rarely exercised at the microarchitectural level, the individual registers and logic cones that implement it at the gate level will likewise experience rare switching activity.~\covert~prunes the rare microarchitectural event space for a given ISA based on publicly available information and open-source implementations and develops tests to thoroughly explore the potential Trojan design space based on the events.    
The key contributions of this paper include: 
\begin{itemize}
    \item We propose a novel framework, \covert, that enables Trojan detection in COTS microprocessors without access to their implementation by transferring activation knowledge from ISA and existing processor designs.
    \item We present a systematic process for extracting rare microarchitectural events from known RTL implementations of microprocessors. To activate these events, we have developed an LLM-assisted program generation framework that consists of an agent-based workflow in LangGraph (built on LangChain)  with iterative mitigation of program errors through automated feedback.
    \item We have extensively evaluated \covert~by applying test programs generated from the OpenRISC mor1kx Cappuccino reference design to two black-box COTS processors: OpenRISC Marocchino, which shares the same ISA, and PicoRV32, which follows a similar RISC design philosophy. We observed over 80\% trigger coverage for the rarest event groups in both. We also analyzed coverage for complex Trojans with multiple excitation of these events. We have open-sourced all the related artifacts from our experiments in this link \cite{artifact}.
\end{itemize}

\section{Background}

\subsection{Trust Issues in COTS Microprocessors}

The design process of a custom IC goes through several steps starting from the specification to all the way to the layout, as shown in Fig.~\ref{fig:design_flow}(a).
Alternatively, if a COTS IC is available in the market, a system developer can avoid these steps and directly obtain a COTS component from the market. For example, a COTS processor based on desired specifications, performance, and ISA can be procured, as shown in Fig.~\ref{fig:design_flow}(b). 
However, the use of COTS hardware introduces trust issues for the procurer due to limited visibility into the design and manufacturing processes, where any of these entities can introduce malicious modification or hardware Trojans.

\subsection{Existing Verification Techniques}
Existing hardware Trojan detection methods are applicable under one or more of the assumptions: i) access to a golden (Trojan-free) design as reference, ii) white-box accessibility to the chip design, iii) design modification to facilitate verification. None of these requirements can, however, be met in the context of COTS components \cite{jain2021survey,cruz2025survey, xiao2016hardware,variant_cots}. For instance, logic testing methods such as MERO \cite{chakraborty2009mero} aim to generate test vectors to ensure activation of hard-to-activate rare nets within the design.  However, existing test generation methods focusing on rare net activation rely on access to the netlist \cite{chakraborty2009mero,8353873,latent}. A few existing verification techniques leverage side channels (e.g., power or delay). Authors in \cite{stern2020sparta} employ a laser-assisted side channel to observe suspicious hard-to-activate flip-flops that are not part of the scan chain. In \cite{yang2021trusted}, the authors propose unsupervised clustering of power side-channels to identify suspicious chips without a golden signature. Both studies, however, require awareness of the design information to create test patterns that activate the rare nets to their rare values.

\subsection{LLM-assisted Processor Verification}

LLM-based test program generation methods have been developed for functional verification and bug detection \cite{llm_processor_1, llm_processor_2, llm_survey, llm_processor_3}. These techniques have shown potential in achieving high code/functional coverage without access to the processor design. 
However, they target functional verification for uncovering unintentional design bugs rather than trust verification, and they do not generate programs that exercise complex microarchitectural events or activate the rare nets that often serve as Trojan triggers. They also cannot be readily extended to Trojan detection because they lack a security-oriented coverage plan, they assume a golden model for correctness, they do not redefine coverage bins around security targets, and they require access to internal RTL coverage in some cases, which is infeasible for COTS. Table \ref{tab:background} provides a comparative analysis of the proposed \covert~framework against existing verification techniques applicable to processors, highlighting their limitations in detecting HTs in COTS hardware.

\begin{figure*}[!htbp]
    \centering
    \includegraphics[width=0.75\textwidth]{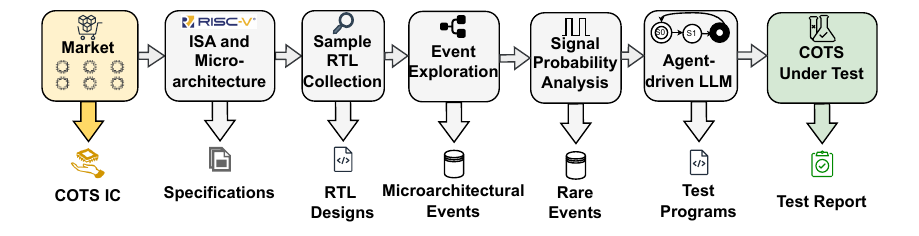}
     \caption{\covert~framework overview for Trojan detection in COTS hardware.}
    \label{fig:overview}
\end{figure*}

\begin{table*}[t]
\centering
\caption{Comparative analysis of existing verification techniques vs.\ the proposed \covert~framework.}
\label{tab:background}
\resizebox{0.75\textwidth}{!}{%
\begin{tabular}{|c|c|c|c|c|}
\hline
\textbf{Proposed Solution} & \textbf{Goal} & \textbf{Task} & \textbf{Trojan Det.?} & \textbf{\#events / \#prog.} \\ \hline
LLM4DV \cite{llm_processor_1}    & Functional Verification            & Stimulus generation       & No  & N/A / N/A  \\ \hline
Xiao et al. \cite{llm_processor_2} & Processor Verification             & Test program generation    & No  & N/A / 164  \\ \hline
LLM-TG  \cite{llm_processor_3}   & Processor Verification             & Test generation            & No  & N/A / 132  \\ \hline
Paria et al. \cite{ip_cots}      & COTS Verification                  & Test program generation    & No  & 9--11 / $\sim$10 \\ \hline
\covert~(this work)             & COTS Verification + Trojan Detection & Test program generation    & Yes & 368 / $\sim$500 \\ \hline
\end{tabular}%
}
\end{table*}

\section{Methodology}

Fig.~\ref{fig:overview} illustrates the proposed \covert~framework. 
The process begins by identifying the ISA of the target COTS microprocessor and obtaining an open-source RTL design that implements the same ISA or follows the same architecture philosophy. From this RTL, we extract detailed microarchitectural information to define high-level events. Signal-probability analysis is then used to rank these events by rarity. Finally, a state-driven agent generates test programs to trigger the rare events and reports their trigger coverage.

\subsection{Microarchitectural Event Explorartion}
\label{ss_aei}

\begin{figure*}[t!]
    \centering
    \includegraphics[width=0.75\textwidth]{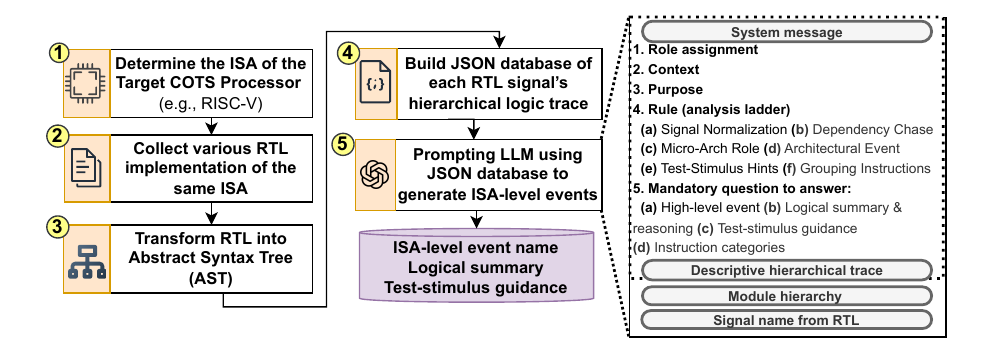}
     \caption{Overview of the event exploration process, where it maps RTL signals into microarchitectural events by hierarchical logic unrolling through AST.}
    \label{fig:ISA_event}
\end{figure*}

A \textit{microarchitectural event} refers to an observable internal condition or transition in the processor’s microarchitecture that encapsulates a meaningful high-level behavior resulting from the activation of specific low-level nets within the microprocessor. For example, if a rare net is influenced by the instruction page-fault flag, privilege mode bits, and TLB miss status, it can be abstracted as the microarchitectural event: ``instruction page fault in user mode with pending TLB refill.'' Such event abstractions effectively map low-level signals to architecturally invariant state, enabling precise targeting through test programs. Here, \textbf{architecturally invariant means the event remains valid and maintains identical semantics across processors that implement the same ISA, regardless of internal design}.
The systematic flow of event exploration is illustrated in Fig.~\ref{fig:ISA_event}.

The process begins with \circled{1} \textbf{identifying the ISA of the target COTS microprocessor} and selecting representative RTL implementations that conform to the same ISA and specification. One \circled{2} \textbf{Verilog RTL design} is chosen as the \textbf{reference} implementation to facilitate rare event identification and subsequent targeted test generation. \textcolor{black}{We chose the RTL design for event generation because it exposes the internal control logic and dependencies needed to locate hard-to-reach conditions that public ISA manuals and datasheets do not reveal.}
The selected \circled{3} \textbf{RTL is then parsed and transformed into an AST} 
enabling both structural and semantic analysis of the design. A custom script \circled{4} \textbf{recursively traces each identified rare signal} backward through its fan-in logic in the AST, following both intra-module and cross-module signal dependencies up to a predefined depth. This backward tracing is typically bounded at the decoder stage, which is ideal for revealing the specific instructions responsible for activating the rare signal. The resulting hierarchical trace provides a crucial link between low-level net activity and corresponding high-level architectural behavior, stored in a JSON database. The trace is further abstracted into simplified textual representations of procedural and control logic blocks, providing the necessary \circled{5} \textbf{context for the LLM to interpret structural dependencies} in natural language and \textbf{infer corresponding microarchitectural events}. To maintain semantic accuracy and prevent hallucinations, the LLM is driven by a carefully crafted system prompt specifying its role, contextual information, analysis rules, and the expected output format. Based on this, the LLM generates event names with high-level descriptions linked to the traced signals, along with architectural summaries, test generation guidance, and relevant instruction categories, serving as the foundation for producing targeted test programs.

\subsection{Rare Event Identification}
\label{rare_event_iden}

We perform rarity analysis to identify microarchitectural events that occur infrequently and serve as Trojan triggers. The reference RTL is stimulated with diverse benchmark programs that generate realistic switching activity. 
These benchmarks are chosen to cover a wide range of instruction types, data patterns, and control-flow behaviors, providing a realistic view of real-world applications and enabling extraction of rare events. During simulation, Value Change Dump (VCD) traces are generated, capturing signal transitions over time. 
From the VCD text files, Signals are classified into \textit{single-bit nets} and \textit{multi-bit buses}, 
each requiring a distinct activity metric for rarity analysis. 
For \textbf{single-bit nets}, the probabilities of the signal being at logic high
($\hat{p}_1$) and logic low ($\hat{p}_0$) are estimated over the simulation cycles.
The rarity metric is defined as $\theta = \min(\hat{p}_0,\hat{p}_1)$.
For \textbf{multi-bit buses}, activity is measured using the normalized toggle rate
defined as:

\[
\theta= \frac{\text{Number of value changes}}{\text{Total number of consecutive cycle pairs}}
\]

\begin{figure}[!htbp]
    \centering
    \includegraphics[width=\columnwidth]{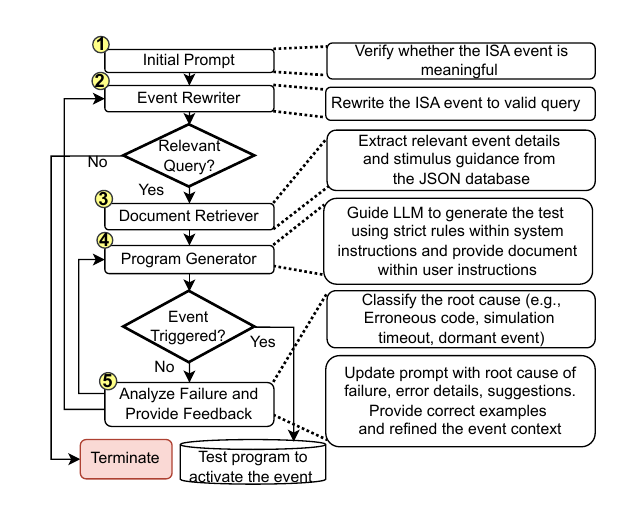}
    \vspace{-1em}
     \caption{Proposed agent-driven workflow in \covert~for test program generation leveraging LLM to activate microarchitectural rare events.}
    \label{fig:llm_agent}

\end{figure}

\subsection{Test Program Generation by leveraging LLM}

\covert~integrates the automated software code generation capabilities of LLMs to generate targeted test programs in C/C++ or inline assembly. 
An LLM-based state-driven agentic workflow, where each state performs a specific, sequenced task, for program generation is illustrated in Fig.~\ref{fig:llm_agent}. The following subsections describe each state briefly.

\subsubsection{Event Rewriter and Topic Classifier}

\covert~starts with processing the user-provided event query to \circled{1} validate its correctness, ensuring the event name is valid, consistent, and free from linguistic errors. If errors are detected, the agent \circled{2} rewrites the query while preserving its intended semantics. The revised query is then passed to a topic classifier, which verifies whether it maps to a valid microarchitectural event under the constraints of the target ISA.

\subsubsection{Document Retriever}\label{DR_ss}
After validating the event query, the agent proceeds to \circled{3} retrieve microarchitectural knowledge associated with the event. This step is critical for guiding the LLM to generate meaningful programs beyond surface-level event names by providing the underlying logical reasoning. The knowledge base for this retrieval is constructed during the event generation phase (Section \ref{ss_aei}) and stored in a structured JSON file. For each event, the database contains three key metadata fields in addition to the event name: \textbf{(a)} a logical summary describing the reasoning chain that leads to the event, \textbf{(b)} stimulus-generation guidance, and \textbf{(c)} relevant instruction categories indicating which ISA operations are most applicable.

\begin{figure*}[!htbp]
    \centering
    \includegraphics[width=0.9\textwidth]{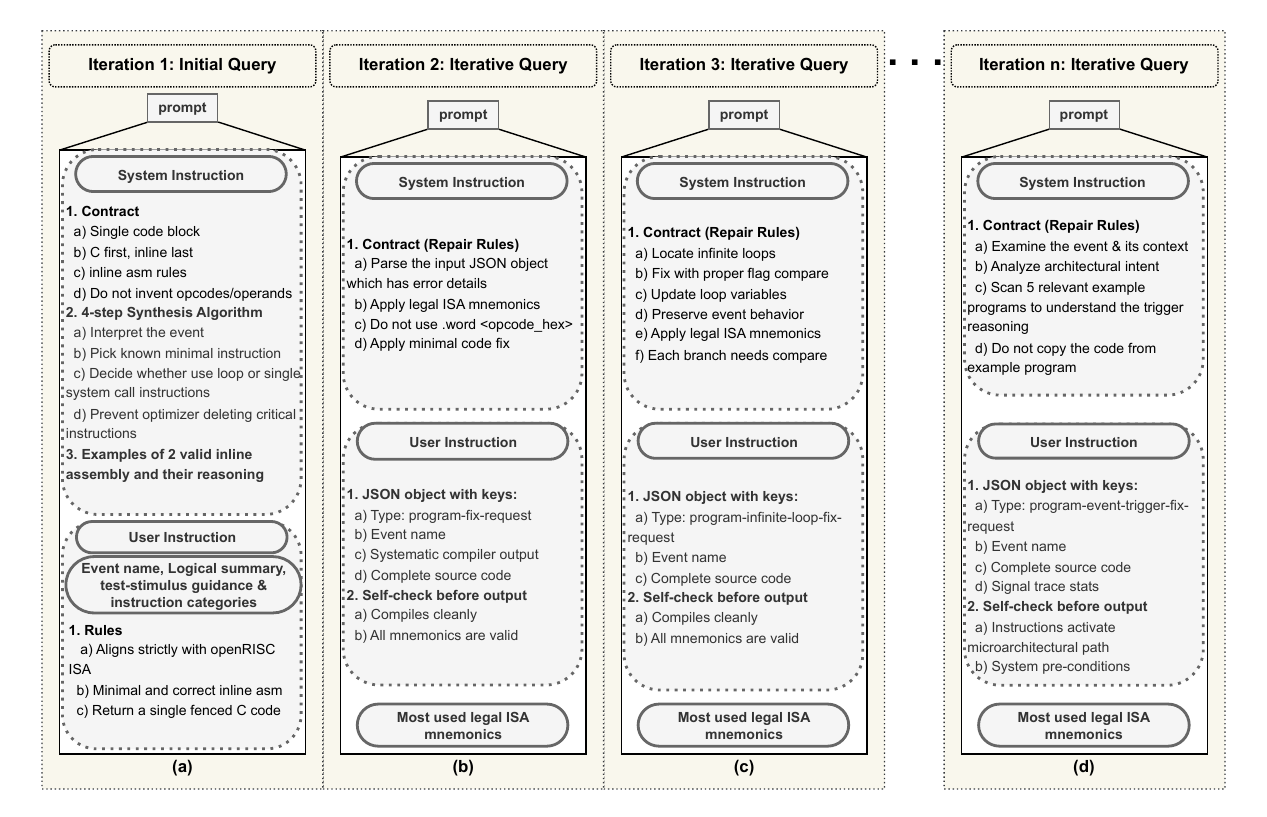}
    \vspace{-1.25em}
     \caption{Overview of iterative prompt refinement process in \covert~framework. The figure highlights the progression from (a) initial query through (b-c) repair-based iterations and (d) subsequent refinements. The initial prompt instructs the LLM to generate a minimal C/C++ or inline assembly program that satisfies the event's conditions. If compilation or simulation fails, the error output is appended to the next prompt so the model can repair the code. If the program runs but does not activate the event, logical feedback and previous successful examples are added to the next prompt. This loop continues until the refined prompt produces a legal program that successfully triggers the rare event.}
    \label{fig:query}
\end{figure*}

\subsubsection{Program Generator}

\covert~leverages the LLM to \circled{4} generate test programs using a systematic prompt template. The process begins with an iteration check, with no feedback applied in the first iteration due to the absence of prior output. The state then iterates until the generated program meets the trigger condition. If the generated program fails, the agent refines the prompt by incorporating failure feedback and tightening constraints, progressively increasing the likelihood of success. The prompt is organized using modern LLM APIs into two layers: system and user instructions, allowing dynamic updates to the user layer while keeping the system layer tasks stable.
The system prompt defines the model’s role, enforces global rules (e.g., single code block, no fabricated opcodes), and outlines a step-by-step analysis procedure. It also includes k-shot learning through working examples: one in each standard C and inline assembly, to demonstrate the expected output format. The user prompt includes four key components: (a) event-specific metadata from previous state (see Section \ref{DR_ss}), (b) user-defined constraints and objectives, (c) a pre-generation self-checklist, and (d) a strictly defined output structure.

Another challenge is that LLMs often struggle to consistently extract correct ISA semantics from large, unstructured PDF manuals. To mitigate this, the ISA documentation is restructured into two parts: (a) JSON representation capturing valid opcodes, operands, and encoding rules; and (b) the original PDF content for high-level architectural context (e.g., memory models, exceptions, and special-purpose registers). This dual-format knowledge base, combined with an automated retrieval-augmentation mechanism, significantly improves code correctness and validity and also reduces the number of required iterations for successful event triggering.

\subsubsection{Detection of Event Trigger}
In this state, the agent compiles and simulates the test program generated in the previous stage and produces the VCD trace. This trace is then compared against a golden reference trace, generated by executing an empty main function, to detect microarchitectural transition differences. An increase in transitions indicates that the test program successfully triggered the targeted event and reached the termination state. If the event is not triggered, the framework attributes the failure to one of three primary causes: (a) compilation error, (b) simulation timeout, or (c) absence of event activation. The agent logs the failure reason and associated outputs, which are then forwarded to the feedback stage to guide the next iteration of test generation.

\subsubsection{Failure Analyzer and Feedback}

The primary objective is to establish a structured \circled{5} feedback loop that enables LLM to iteratively refine its output based on the root causes of failure. The agent analyzes failed outputs from one of three primary causes and incorporates targeted feedback into subsequent prompt iterations. For compiler errors, the agent parses the compiler log into a structured JSON format detailing the error type, location, and cause. For simulation timeouts, the system prompt is updated to instruct the LLM to detect potential infinite loops, add termination conditions, or revise problematic inline assembly segments. When a program compiles and executes but fails to trigger the target event, the agent supplements the prompt with previously successful, event-relevant examples to improve context. Throughout all iterations, the system prompt enforces repair strategies and contractual rules, while the user prompt includes structured error reports and a history of prior code attempts. 
To aid in generating valid inline assembly, the agent adds a list of frequently used ISA instructions to the user prompt, reducing the model’s need to search the JSON database for targeted instructions.
This guided feedback mechanism ensures that LLM receives precise, context-aware information, significantly improving convergence toward valid and effective test programs. 
Fig. \ref{fig:query} demonstrates the iterative prompt refinement process for LLM in \covert~framework.

\section{Results}
\label{results}

We have used the mor1kx Cappuccino core\cite{openrisc_mor1kx} from OpenRISC\cite{openrisc_project_overview} architecture as the sample design to extract microarchitectural events. For verification, we selected the or1k Marocchino\cite{openrisc_or1k_marocchino} processor, which implements the exact ISA but follows an out-of-order pipeline, and the PicoRV32\cite{yosyshq_picorv32} processor, which follows a similar ISA and design philosophy. 
We applied the CHStone\cite{chstone} and MiBench\cite{mibench} benchmark suites under random testing using Verilator\cite{verilator} to obtain cycle-accurate simulations, then computed signal probabilities to identify rare nets.
PyVerilog\cite{pyverilog} was used to parse the RTL and generate an AST for mapping these rare nets to architectural events.
We completed the framework using LangGraph from LangChain\cite{langgraph} to build a state-driven agent for automated test generation.
The agent connected to the OpenAI Assistant API\cite{openai_assistant}, using its file assistant to retrieve ISA documentation information. Inside the API, we selected the OpenAI 4.1 model (mini), which provided strong coding ability and offered better cost-performance efficiency compared to larger models. All experiments were performed on a 24-core Intel Core i7-13700F system (2.1 GHz base, 64 GB RAM) running Ubuntu 22.04.5. We have open-sourced all related artifacts in \cite{artifact}.

\begin{table}[t]
\centering
\begingroup
\caption{Signals distribution across pipeline modules of mor1kx Cappuccino with examples of RTL signals mapped to their high-level events.}
\scriptsize
\setlength{\tabcolsep}{2pt}        
\renewcommand{\arraystretch}{1.05}  
\setlength{\arrayrulewidth}{0.5pt} 
\fontsize{7}{8}\selectfont

\begin{tabularx}{\columnwidth}{|l|
  >{\centering\arraybackslash}p{0.09\columnwidth}|
  >{\centering\arraybackslash}p{0.08\columnwidth}|
  >{\centering\arraybackslash}p{0.08\columnwidth}|
  >{\centering\arraybackslash}p{0.17\columnwidth}|
  Y|Y|Y|}
\hline
\textbf{Stage} & \textbf{Total Signals} & \textbf{Total Events} & \textbf{Rare Events} &
\textbf{Example Signal} & \textbf{Corresponding High-Level Event} \\
\hline
Decode & 56 & 45 & 11 & decode\_op\_movhi\_o &
Pipeline stall on instruction decode due to dependency on a high immediate value. \\
\hline
Fetch & 237 & 149 & 65 & icache\_refill\_done\_o &
Cache refill completes, allowing instruction fetch to resume. \\
\hline
ALU & 93 & 76 & 16 & overflow\_set\_o &
Detection of signed overflow or division by zero. \\
\hline
LSU & 278 & 173 & 37 & align\_err\_short &
Misaligned memory access. \\
\hline
Ctrl & 194 & 90 & 25 & ctrl\_op\_mtspr\_i &
Move-to-SPR instruction transfers data from General Purpose Register (PR) to Special PR.\\
\hline
\end{tabularx}
\label{tab:rare-events}
\endgroup
\end{table}

\subsection{Mapping Signals to Architectural Events}  
Table~\ref{tab:rare-events} summarizes the available signals across the five pipeline modules of the mor1kx Cappuccino core. In principle, microarchitectural events may be derived from meaningful combinations of multiple signals. However, in our experiments, we treat each signal as a potential architectural event, while excluding signals that never toggled during benchmark simulation or failed during AST abstraction. We also reported the number of rare events when rarity threshold $\theta$<=0.05 and $\theta$$\in${[}0,1{]}.
Table~\ref{tab:rare-events} also illustrates five representative examples, one from each pipeline module, where RTL signals are mapped to architectural events.

\subsection{Test Program Generation for COTS}

\begin{figure}[t!]
    \centering
    \includegraphics[width=1\columnwidth]{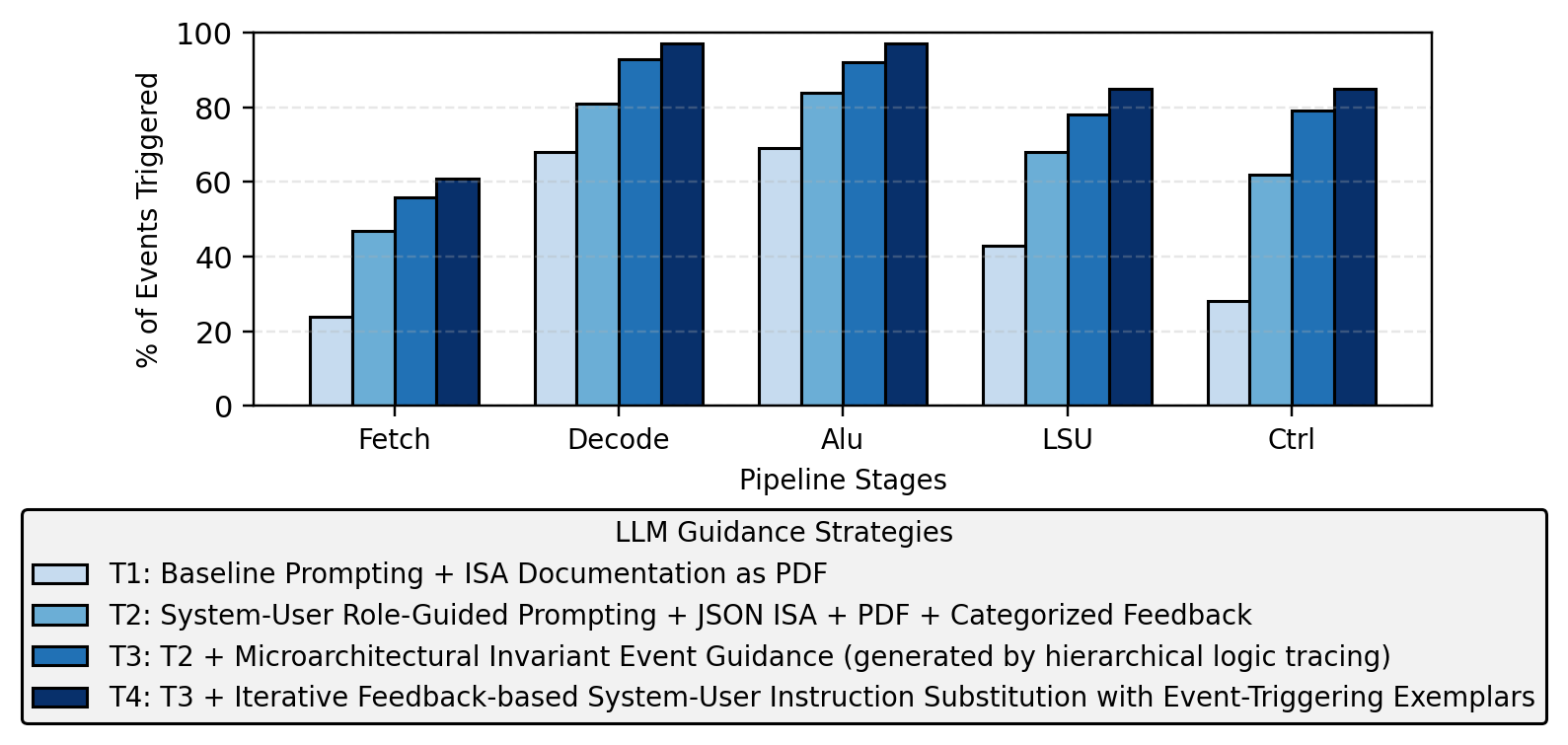}
     \vspace{-2em}
     \caption{Percentage of events triggered in sample design across pipeline modules using four progressively refined prompting strategies (T1–T4).}
    \label{fig:triggered_percentage}
\end{figure}

To produce valid high-level and bare-metal programs by modern LLMs that reliably trigger specific events requires precise knowledge of register addresses and bit-level configurations, which an LLM cannot infer from ISA documentation.
To overcome these limitations, we progressively refined our prompting strategies.  Fig.~\ref{fig:triggered_percentage} shows how each refinement, starting from baseline prompting to iterative feedback, increased the percentage of events triggered in the sample design.

\textbf{T1. Baseline Prompting:} In the initial stage, only the ISA documentation in PDF format was provided to the LLM (GPT 4.1 mini), with minimal rule-setting restricted to the user role. The event name and a brief architectural summary were supplied as input. Under this setting, we observed the lowest rate of correct test program generation. Most outputs either failed to compile or resulted in simulation timeouts, leading to a lower number of events successfully triggered across pipeline modules shown in Fig.~\ref{fig:triggered_percentage}.

\begin{table*}[htb]
\centering
\caption{Evaluation of \covert~using test programs from the sample mor1kx Cappuccino RTL on two targets: module-wise rare-event coverage for the OpenRISC Marocchino processor and overall rare-event coverage for the single-module RISC-V PicoRV32.}
\label{tab:tc_pico}
\resizebox{\textwidth}{!}{%
\begin{tabular}{|cccccccccccccccc|lll|}
\hline
\multicolumn{16}{|c|}{\textbf{OpenRISC Marocchino}} &
  \multicolumn{3}{c|}{\textbf{RISC-V PICORV32IMC}} \\ \hline
\multicolumn{1}{|l|}{\textbf{module}} &
  \multicolumn{3}{c|}{\textbf{Fetch}} &
  \multicolumn{3}{c|}{\textbf{Decode}} &
  \multicolumn{3}{c|}{\textbf{ALU}} &
  \multicolumn{3}{c|}{\textbf{LSU}} &
  \multicolumn{3}{c|}{\textbf{Ctrl}} &
  \multicolumn{3}{c|}{\textbf{Combined}} \\ \hline
\multicolumn{1}{|c|}{\textbf{$\theta$}} &
  \multicolumn{1}{c|}{\textbf{\begin{tabular}[c]{@{}c@{}}Rare\\ events\end{tabular}}} &
  \multicolumn{1}{c|}{\textbf{\begin{tabular}[c]{@{}c@{}}Rare\\ events\\ triggered\end{tabular}}} &
  \multicolumn{1}{c|}{\textbf{\begin{tabular}[c]{@{}c@{}}\%   of\\ triggered\end{tabular}}} &
  \multicolumn{1}{c|}{\textbf{\begin{tabular}[c]{@{}c@{}}Rare\\ events\end{tabular}}} &
  \multicolumn{1}{c|}{\textbf{\begin{tabular}[c]{@{}c@{}}Rare\\ events\\ triggered\end{tabular}}} &
  \multicolumn{1}{c|}{\textbf{\begin{tabular}[c]{@{}c@{}}\% of\\ triggered\end{tabular}}} &
  \multicolumn{1}{c|}{\textbf{\begin{tabular}[c]{@{}c@{}}Rare\\ events\end{tabular}}} &
  \multicolumn{1}{c|}{\textbf{\begin{tabular}[c]{@{}c@{}}Rare\\ events\\ triggered\end{tabular}}} &
  \multicolumn{1}{c|}{\textbf{\begin{tabular}[c]{@{}c@{}}\%   of\\ triggered\end{tabular}}} &
  \multicolumn{1}{c|}{\textbf{\begin{tabular}[c]{@{}c@{}}Rare\\ events\end{tabular}}} &
  \multicolumn{1}{c|}{\textbf{\begin{tabular}[c]{@{}c@{}}Rare\\ events\\ triggered\end{tabular}}} &
  \multicolumn{1}{c|}{\textbf{\begin{tabular}[c]{@{}c@{}}\%   of \\ triggered\end{tabular}}} &
  \multicolumn{1}{c|}{\textbf{\begin{tabular}[c]{@{}c@{}}Rare \\ events\end{tabular}}} &
  \multicolumn{1}{c|}{\textbf{\begin{tabular}[c]{@{}c@{}}Rare\\ events \\ triggered\end{tabular}}} &
  \textbf{\begin{tabular}[c]{@{}c@{}}\%   of\\ triggered\end{tabular}} &
  \multicolumn{1}{c|}{\textbf{\begin{tabular}[c]{@{}c@{}}Rare\\ events\end{tabular}}} &
  \multicolumn{1}{c|}{\textbf{\begin{tabular}[c]{@{}c@{}}Rare\\ events\\ triggered\end{tabular}}} &
  \multicolumn{1}{c|}{\textbf{\begin{tabular}[c]{@{}c@{}}\%   of \\ triggered\end{tabular}}} \\ \cline{2-19} 
\multicolumn{1}{|c|}{} &
  \multicolumn{3}{c|}{\textbf{Test programs simulated = 80}} &
  \multicolumn{3}{c|}{\textbf{Test programs simulated = 41}} &
  \multicolumn{3}{c|}{\textbf{Test programs simulated = 70}} &
  \multicolumn{3}{c|}{\textbf{Test programs simulated = 91}} &
  \multicolumn{3}{c|}{\textbf{Test programs simulated =77}} &
  \multicolumn{3}{c|}{\textbf{Test programs simulated =364}} \\ \hline
\multicolumn{1}{|c|}{0.05} &
  \multicolumn{1}{c|}{78} &
  \multicolumn{1}{c|}{62} &
  \multicolumn{1}{c|}{79.49} &
  \multicolumn{1}{c|}{21} &
  \multicolumn{1}{c|}{21} &
  \multicolumn{1}{c|}{100} &
  \multicolumn{1}{c|}{12} &
  \multicolumn{1}{c|}{11} &
  \multicolumn{1}{c|}{91.67} &
  \multicolumn{1}{c|}{107} &
  \multicolumn{1}{c|}{85} &
  \multicolumn{1}{c|}{79.44} &
  \multicolumn{1}{c|}{86} &
  \multicolumn{1}{c|}{74} &
  86.05 &
  \multicolumn{1}{l|}{101} &
  \multicolumn{1}{l|}{86} &
  85.15 \\ \hline
\multicolumn{1}{|c|}{0.15} &
  \multicolumn{1}{c|}{110} &
  \multicolumn{1}{c|}{94} &
  \multicolumn{1}{c|}{85.45} &
  \multicolumn{1}{c|}{34} &
  \multicolumn{1}{c|}{34} &
  \multicolumn{1}{c|}{100} &
  \multicolumn{1}{c|}{26} &
  \multicolumn{1}{c|}{25} &
  \multicolumn{1}{c|}{96.15} &
  \multicolumn{1}{c|}{195} &
  \multicolumn{1}{c|}{172} &
  \multicolumn{1}{c|}{88.21} &
  \multicolumn{1}{c|}{95} &
  \multicolumn{1}{c|}{83} &
  87.37 &
  \multicolumn{1}{l|}{134} &
  \multicolumn{1}{l|}{119} &
  88.81 \\ \hline
\multicolumn{1}{|c|}{0.25} &
  \multicolumn{1}{c|}{120} &
  \multicolumn{1}{c|}{102} &
  \multicolumn{1}{c|}{85.00} &
  \multicolumn{1}{c|}{41} &
  \multicolumn{1}{c|}{41} &
  \multicolumn{1}{c|}{100} &
  \multicolumn{1}{c|}{42} &
  \multicolumn{1}{c|}{41} &
  \multicolumn{1}{c|}{97.62} &
  \multicolumn{1}{c|}{220} &
  \multicolumn{1}{c|}{197} &
  \multicolumn{1}{c|}{89.55} &
  \multicolumn{1}{c|}{99} &
  \multicolumn{1}{c|}{87} &
  87.88 &
  \multicolumn{1}{l|}{147} &
  \multicolumn{1}{l|}{132} &
  89.80 \\ \hline
\multicolumn{1}{|c|}{0.35} &
  \multicolumn{1}{c|}{125} &
  \multicolumn{1}{c|}{106} &
  \multicolumn{1}{c|}{84.80} &
  \multicolumn{1}{c|}{50} &
  \multicolumn{1}{c|}{50} &
  \multicolumn{1}{c|}{100} &
  \multicolumn{1}{c|}{56} &
  \multicolumn{1}{c|}{55} &
  \multicolumn{1}{c|}{98.21} &
  \multicolumn{1}{c|}{247} &
  \multicolumn{1}{c|}{224} &
  \multicolumn{1}{c|}{90.69} &
  \multicolumn{1}{c|}{103} &
  \multicolumn{1}{c|}{91} &
  88.35 &
  \multicolumn{1}{l|}{154} &
  \multicolumn{1}{l|}{139} &
  90.26 \\ \hline
\multicolumn{1}{|c|}{0.45} &
  \multicolumn{1}{c|}{130} &
  \multicolumn{1}{c|}{111} &
  \multicolumn{1}{c|}{85.38} &
  \multicolumn{1}{c|}{71} &
  \multicolumn{1}{c|}{71} &
  \multicolumn{1}{c|}{100} &
  \multicolumn{1}{c|}{63} &
  \multicolumn{1}{c|}{62} &
  \multicolumn{1}{c|}{98.41} &
  \multicolumn{1}{c|}{253} &
  \multicolumn{1}{c|}{230} &
  \multicolumn{1}{c|}{90.91} &
  \multicolumn{1}{c|}{108} &
  \multicolumn{1}{c|}{96} &
  88.89 &
  \multicolumn{1}{l|}{167} &
  \multicolumn{1}{l|}{152} &
  91.02 \\ \hline
\multicolumn{1}{|c|}{0.55} &
  \multicolumn{1}{c|}{165} &
  \multicolumn{1}{c|}{143} &
  \multicolumn{1}{c|}{86.67} &
  \multicolumn{1}{c|}{90} &
  \multicolumn{1}{c|}{90} &
  \multicolumn{1}{c|}{100} &
  \multicolumn{1}{c|}{65} &
  \multicolumn{1}{c|}{64} &
  \multicolumn{1}{c|}{98.46} &
  \multicolumn{1}{c|}{258} &
  \multicolumn{1}{c|}{235} &
  \multicolumn{1}{c|}{91.09} &
  \multicolumn{1}{c|}{116} &
  \multicolumn{1}{c|}{104} &
  89.66 &
  \multicolumn{1}{l|}{170} &
  \multicolumn{1}{l|}{155} &
  91.18 \\ \hline
\multicolumn{1}{|c|}{0.65} &
  \multicolumn{1}{c|}{176} &
  \multicolumn{1}{c|}{153} &
  \multicolumn{1}{c|}{86.93} &
  \multicolumn{1}{c|}{94} &
  \multicolumn{1}{c|}{94} &
  \multicolumn{1}{c|}{100} &
  \multicolumn{1}{c|}{67} &
  \multicolumn{1}{c|}{66} &
  \multicolumn{1}{c|}{98.51} &
  \multicolumn{1}{c|}{260} &
  \multicolumn{1}{c|}{237} &
  \multicolumn{1}{c|}{91.15} &
  \multicolumn{1}{c|}{120} &
  \multicolumn{1}{c|}{108} &
  90.00 &
  \multicolumn{1}{l|}{176} &
  \multicolumn{1}{l|}{161} &
  91.48 \\ \hline
\multicolumn{1}{|c|}{0.75} &
  \multicolumn{1}{c|}{180} &
  \multicolumn{1}{c|}{157} &
  \multicolumn{1}{c|}{87.22} &
  \multicolumn{1}{c|}{99} &
  \multicolumn{1}{c|}{99} &
  \multicolumn{1}{c|}{100} &
  \multicolumn{1}{c|}{70} &
  \multicolumn{1}{c|}{69} &
  \multicolumn{1}{c|}{98.57} &
  \multicolumn{1}{c|}{260} &
  \multicolumn{1}{c|}{237} &
  \multicolumn{1}{c|}{91.15} &
  \multicolumn{1}{c|}{120} &
  \multicolumn{1}{c|}{108} &
  90.00 &
  \multicolumn{1}{l|}{179} &
  \multicolumn{1}{l|}{164} &
  91.62 \\ \hline
\multicolumn{1}{|c|}{0.85} &
  \multicolumn{1}{c|}{183} &
  \multicolumn{1}{c|}{160} &
  \multicolumn{1}{c|}{87.43} &
  \multicolumn{1}{c|}{101} &
  \multicolumn{1}{c|}{101} &
  \multicolumn{1}{c|}{100} &
  \multicolumn{1}{c|}{73} &
  \multicolumn{1}{c|}{72} &
  \multicolumn{1}{c|}{98.63} &
  \multicolumn{1}{c|}{261} &
  \multicolumn{1}{c|}{238} &
  \multicolumn{1}{c|}{91.19} &
  \multicolumn{1}{c|}{125} &
  \multicolumn{1}{c|}{113} &
  90.40 &
  \multicolumn{1}{l|}{179} &
  \multicolumn{1}{l|}{164} &
  91.62 \\ \hline
\multicolumn{1}{|c|}{1} &
  \multicolumn{1}{c|}{199} &
  \multicolumn{1}{c|}{175} &
  \multicolumn{1}{c|}{87.94} &
  \multicolumn{1}{c|}{104} &
  \multicolumn{1}{c|}{104} &
  \multicolumn{1}{c|}{100} &
  \multicolumn{1}{c|}{75} &
  \multicolumn{1}{c|}{74} &
  \multicolumn{1}{c|}{98.67} &
  \multicolumn{1}{c|}{266} &
  \multicolumn{1}{c|}{243} &
  \multicolumn{1}{c|}{91.35} &
  \multicolumn{1}{c|}{135} &
  \multicolumn{1}{c|}{123} &
  91.11 &
  \multicolumn{1}{l|}{182} &
  \multicolumn{1}{l|}{167} &
  91.76 \\ \hline
\end{tabular}%
}
\end{table*}

\textbf{T2. Role-Guided Prompting:} 
After identifying the common issues, we addressed them in two approaches: (i) we converted the ISA’s table-based descriptions from the PDF into a JSON schema 
\{\texttt{id}, \texttt{syntax}, \texttt{encoding}, \texttt{example\_hex}, \texttt{description}\}, and retaining the PDF only for memory and Special Purpose Register (SPR) details; 
(ii) we enforced strict coding rules and a repair loop that feeds the compiler and simulation errors back to the LLM to fix the current program rather than regenerate. We also incorporated compiler header definitions that provide predefined macros and access functions for SPRs. As a result, we were able to reduce errors, which mostly came from invalid opcode/operand choices and SPR bit configurations. Fig.~\ref{fig:triggered_percentage} shows that the number of events triggered increased significantly: Fetch from 24\% to 47\%, LSU from 43\% to 68\% and Ctrl from 28\% to 62\%, while Decode and ALU improved slightly since most of their events already required less low-level programming effort.

\textbf{T3. Incorporating Event Guidance:} For events that still failed to trigger, we added higher-level context, i.e., logical summary of the signal’s behavior, guidance on suitable test stimuli, and categories of instructions likely to influence the event. This abstraction allowed the LLM to connect low-level activity with ISA-level outcomes while remaining architecture-invariant. Building on T2, these additions improved the percentage of events triggered across all pipeline modules.

\textbf{T4. Per Run Role Definition with Exemplars:} 
Finally, T3 was extended with iterative feedback loops where system and user instructions were adjusted based on the current state of program generation. When an event failed to trigger, the framework retrieved a small set of previously successful programs (e.g., 3–5) from the existing pool and supplied them as exemplars alongside the updated instructions. This approach led to further improvements in event triggering, shown in Fig.~\ref {fig:triggered_percentage}. \textcolor{black}{Even after applying T4, coverage stayed low for the fetch module because the LLM generated small, loop-based tests that tried to trigger memory events, such as instruction cache misses and refills, page faults, cache invalidations via special registers, and instruction bus fetches. These tests did not create the necessary setup, including large and varied code, many jumps across far-apart code locations, and a long run time that changes cache and memory settings.
}

\subsection{Event Coverage for COTS}
\label{Event_cov_COTS}
To evaluate the effectiveness of \covert~on unknown COTS processors, we executed the test programs generated from the sample implementation. For COTS, direct one-to-one matching of events generated from signals with the sample design is challenging because of differences in implementation.
Therefore, we first reran the same CHStone and MiBench benchmarks on Marocchino and PicoRV32 to identify rare events. For Marocchino, we simulated module-specific test programs. For PicoRV32, which is organized as a single top-level module without distinct pipeline stages, we executed the entire set of test programs. OpenRISC inline assembly was automatically converted to the RV32IMC instruction set for PicoRV32 using the LLM. 

Table \ref{tab:tc_pico} summarizes the results. With a rarity threshold (see Section \ref{rare_event_iden}) of 5\%, Marocchino achieved nearly 80\% event coverage in the Fetch and LSU modules, while Decode and ALU exceeded 90\%, and Ctrl reached over 86\%. PicoRV32 showed similar activation rates. As the rarity threshold was relaxed, coverage increased steadily across all modules. Decode and ALU depend less on microarchitectural features, whereas Fetch and LSU involve cache, MMU, and SPR interactions, explaining their slightly lower coverage in Marocchino. PicoRV32, a size-optimized core lacking caches, memory management units, and access mode, gained only about 2\% additional coverage even after running the converted assembly programs. \textit{\textbf{These results demonstrate that microarchitecturally derived programs from the sample core design remain effective across different implementations, providing high coverage despite the absence of internal design knowledge.}} COTS microprocessors that follow a similar ISA will always exhibit overlapping rare events, while unique events arise from implementation specific features. By continually generating microarchitectural events from diverse open-source RTL designs, a procurer can progressively cover both the shared and unique rare events, increase rare-node activation, and trust.
A Trojan trigger can be combinational or sequential. By definition, a combinational Trojan is triggered by a single event; therefore, this analysis can be considered combinational trigger coverage.

\subsection{Complex Trigger Coverage for COTS Marocchino}
\begin{figure}[!htbp]
    \centering
    \includegraphics[width=1\columnwidth]{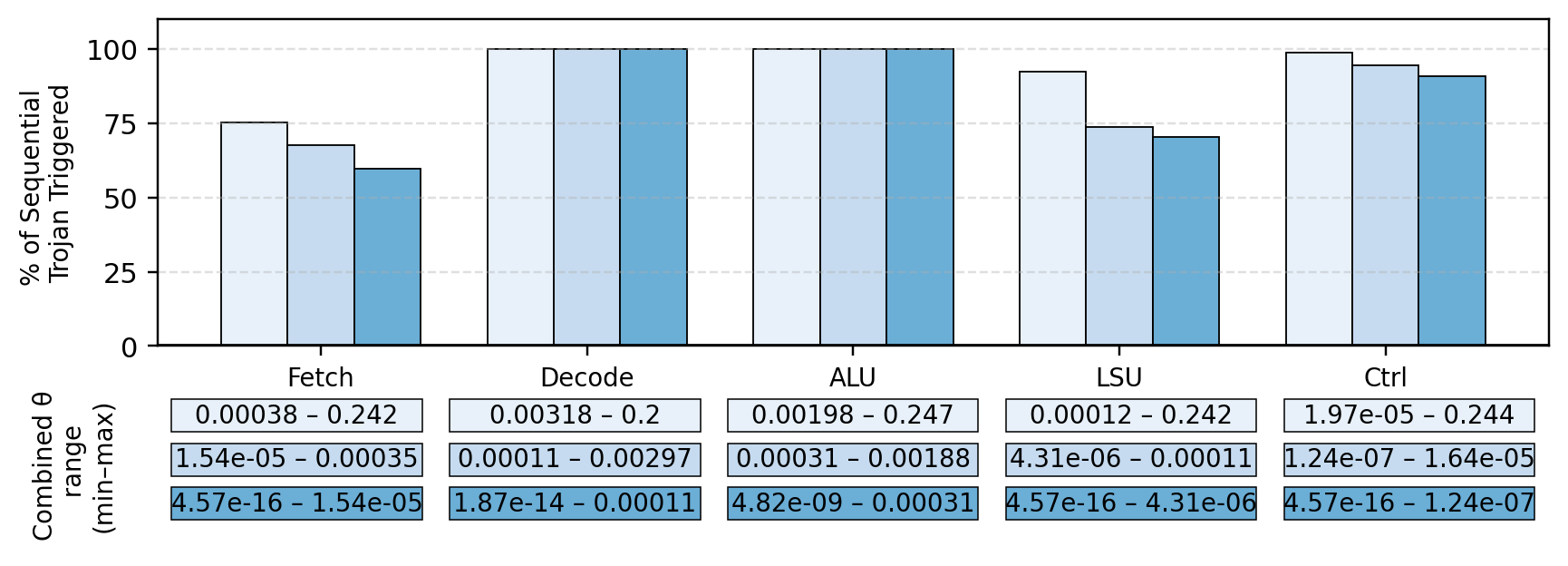}
    \vspace{-2em}
     \caption{ Percentage of sequential trigger coverage across pipeline modules with joint rarity threshold $\theta$ in the OpenRISC Marocchino microprocessor.}
    \label{fig:seq_trojan}
\end{figure}

While an occurrence of one or more distinct events could be considered as combinational triggers, a sequential trigger could be designed using a sequence of multiple events. We extended the analysis by constructing sequential Trojan instances through joint activation of individual events. 
To keep activation probability low but realistic \cite{trusthub_bench}, we keep event combinations within a realistic rarity threshold range shown in Fig.~\ref{fig:seq_trojan}.
We observed that sequential trigger coverage across all modules of the OpenRISC Marocchino ranged from about 60\% in Fetch to nearly 100\% in Decode and ALU, with combined joint $\theta$ values spanning approximately $4.6\times10^{-16}$ to $0.24$.

\subsection{Mathematical Analysis for Coverage}
\textit{Coverage} is calculated as the fraction of Trojans triggered during testing out of the total number of Trojans.

\circled{i} For individual test programs, each Trojan $j$ is targeted by a single test program $t_j$ that triggers it with probability $p_j$.
Assuming all $T$ test programs are applied independently, the expected number of Trojans triggered is:
\setlength{\belowdisplayskip}{0pt} \setlength{\belowdisplayshortskip}{0pt}
\setlength{\abovedisplayskip}{0pt} \setlength{\abovedisplayshortskip}{0pt}
\begin{equation}
    \footnotesize \mathbb{E}[\text{Triggers}] = \sum_{j=1}^{T} p_j
\end{equation}
\setlength{\belowdisplayskip}{0pt} \setlength{\belowdisplayshortskip}{0pt}
\setlength{\abovedisplayskip}{0pt} \setlength{\abovedisplayshortskip}{0pt}
\begin{equation}
\text{Exp. coverage :} \footnotesize \text{\textbf{ Coverage}}_{indv} = \frac{1}{N} \sum_{j=1}^{N} p_j
\end{equation}

\circled{ii} For combining test programs, the probability that event $e_j$ is triggered by at least one test program is:
\begin{equation}
     \footnotesize P_{\text{trigger}, j} = 1 - \prod_{t_i \in \mathcal{T}_j} (1 - p_{ij})
\end{equation}

The expected number of triggered Trojans across all $N$ events is then:
\begin{equation}
     \footnotesize \mathbb{E}[\text{Triggers}] = \sum_{j=1}^{N} \left(1 - \prod_{t_i \in \mathcal{T}_j} (1 - p_{ij}) \right)
\end{equation}
where,  $\{p_{ij}\}_{i=1}^n$ be a set of independent probabilities in $[0, 1)$.

\setlength{\belowdisplayskip}{0pt} \setlength{\belowdisplayshortskip}{0pt}
\setlength{\abovedisplayskip}{0pt} \setlength{\abovedisplayshortskip}{0pt}
\begin{equation}
\text{Exp. coverage :}  \footnotesize \text{\textbf{ Coverage}}_{comb} = \frac{1}{N} \sum_{j=1}^{N} \left(1 - \prod_{t_i \in \mathcal{T}_j} (1 - p_{ij}) \right)
\end{equation}

\noindent It can be shown that $\footnotesize 1 - \prod_{i=1}^n (1 - p_{ij}) \geq \max_i p_{ij}$
with strict inequality if more than one $p_{ij} > 0$. This implies\textbf{~$\text{Coverage}_{comb} \ge \text{Coverage}_{indv}$}.
All of the above trends were observed in our experiments.

\section{Discussion and Conclusion}

We have presented \covert, a practical solution to verification of COTS hardware trust while being scalable (with respect to design size) and flexible (with respect to microarchitectural variations). The central idea behind \covert, a golden-free COTS trust verification framework, is the use of microarchitectural events, which are implementation invariant, and triggering them multiple times using a set of test programs to statistically maximize the probability of triggering Trojans. To our knowledge, this is the first instance of trust verification of COTS microprocessors, which can be used by Original Equipment Manufacturers (OEMs) to verify the trust of untrusted COTS components before integrating them into systems. 

While \covert~is promising, we note that there are significant opportunities to improve coverage, test efficiency, and scalability. 
It includes (1) accounting for trigger coverage drop even under the same ISA when the target COTS includes unseen implementation features such as memory protection mechanisms or error-correction code, and (2) increased diversity of the test programs in terms of instruction mix. There are several methods to achieve this. For instance, adding a random prefix code to a valid test program generated by \covert~can alter both the instruction mix and execution trace, thereby statistically enhancing the probability of activating a random trigger condition. Our future work will investigate these opportunities.

\section{Acknowledgment}
The authors acknowledge support from the Purdue Center for Secure Microelectronics Ecosystem – CSME\#210205.

\bibliographystyle{IEEEtran}
\bibliography{IEEEabrv,references}

\end{document}